# 跨域解耦的端到端确定性网络传输调度机制

吴斌伟[1]，汪 硕[1,2]，谭炜骞[3,1],
（1. 网络通信与安全紫金山实验室，南京 211111；2. 北京邮电大学 网络与交换国家重点实验室，北京 100876；
3. 东南大学 网络空间安全学院，南京 211189；）

**摘要**: 为了解决时间敏感业务的跨域通信问题，提出了一种跨域解耦的端到端确定性网络传输机制，支持端到端时延有上界的跨域传输。通过在网络边缘部署离散整形器，实现各个网络时钟域的解耦，解决传统确定性网络方案中因跨域时钟同步导致的部署和调度复杂程度高、实现困难的问题。此外，该传输机制还扩大了当前跨域确定性网络技术的适用范围：除了周期确定业务流以外，还可支持非周期确定业务流、不确定业务流等类型业务的跨域端对端服务质量保障。最后，以提升网络吞吐量和降低网络开销为目标，提出了一种基于拍卖模型的在线业务流跨域调度算法。仿真结果表明，该机制可以有效实现跨域端到端确定性传输，并且与现有的跨域确定性机制相比，可以有效降低跨域时延。

**关 键 词**: 确定性网络; 网络演算; 跨域传输; 业务流调度
**中图分类号**: TN915.41 **文献标识码**: A

## Mechanism design for the end-to-end deterministic transmissions with decoupled time domains

**Abstract:** This paper proposes an innovative end-to-end deterministic network mechanism to achieve delay-bounded transmissions across multiple network domains. The proposed mechanism installs discrete shapers at the edge of the network domains, which serves to decouple the clock domains of different networks. Thereby, the challenges associated with cross-domain clock synchronization that are inherent in state-of-the-art deterministic mechanisms are mitigated, e.g., high complexity during the system implementation and the traffic scheduling. Moreover, the proposed mechanism enhances the availability of the deterministic networking, i.e., not only periodic deterministic traffic, but also aperiodic deterministic traffic and stochastic flows are enabled to be served. Furthermore, an auction-based online scheduling algorithm is developed to improve network efficiency and reduce cost. Simulation results show that the proposed mechanism can effectively realize the end-to-end delay-bounded transmission across multiple domains. Meanwhile, the cross-domain latency could also be reduced compared to the existing methods.

**Key words:** deterministic networking; network calculus; across-domain transmission; flow scheduling

## 1 背景与挑战

随着智慧物联网技术的发展，互联网与实体经济深度融合，全球互联网已由消费型向生产型转变。"尽力而为"转发的传统网络难以支撑未来业务对差异性服务、确定性、低时延的需求[1]。为适应全球网络变革的新趋势，解决当前互联网面临的核心技术问题，满足未来业务（如工业互联网、远程手术等）对网络确定性服务质量的需求，确定性网络技术受到了的广泛关注。近年来的理论研究和实践探索已使确定性网络技术框架逐渐成熟，形成了以灵活以太网（Flexible Ethernet, FlexE）、时间敏感网络（Time-Sensitive Networking, TSN）、确定性网络（Deterministic Networking, DetNet）、确定性WiFi（Deterministic WiFi, DetWiFi）、5G确定性网络（5G Deterministic Networking, 5GDN）为核心的技术体系[1]。一些典

型应用场景也被筛选出来，包括工业互联网、算力网络、智慧电网、智能网联汽车、元宇宙等[2]。

注意到，实际部署情境中，数据传输通常涉及多个网域。然而，目前大部分确定性网络技术通常仅关注单一网域内的传输质量保证，包括确定性带宽、有限时延/抖动等。例如，FlexE 和 DetNet 关注大规模骨干网络的传输；TSN 关注局域范围内的确定性传输；5GDN 和 DetWiFi 关注无线接入网络的确定性服务保证。因此，如何实现跨越网域的端到端确定性服务质量保证已成为当前引起关注的重点[3]-[5]。

现有文献主要依赖成熟的确定性技术，通过设计边缘整形机制和分层调度模型，实现跨域的端到端确定性服务质量保障。例如，文献[6]将复杂的工厂车间网络划分为多个 TSN 子网，通过层次化调度及边缘整形，实现多个 TSN 子网互联的确定性端到端传输。文献[4]进一步考虑多个工厂车间域间的远程通信，利用时隙无冲突约束模型，严格限制跨域流量在边缘节点的到达时间，将网络覆盖范围进一步扩展到城域级别。文献[7]-[9]利用高带宽光纤资源或 OTN 网络直接连接多个 TSN 网络，实现了跨广域全局时钟同步，从而控制了端到端时延抖动。注意到，这些工作通常假设 TSN 子网间的长距链路是理想的，即不会出现数据丢包或剧烈的时延抖动[9]。然而，现实中负责长距传输的城域网或骨干网由于"微突发"现象，不能保证常数时延和无损传输，这直接限制了上述方案在实际场景中的实用性。

为此，文献[10]利用 DetNet 技术确保 TSN 子网间的长距离链路传输的确定性和可靠性，并提出了基于时隙映射的端到端确定性网络传输方案。文献[11]通过边缘节点的逐流首跳偏置，在链路和时隙级别上实现全网流量的负载均衡，提升网络资源利用率和吞吐量。文献[3]采用类似的方法，将无线接入网也纳入了考虑范围中，实现涵盖无线传输、有线转发、和业务计算的端到端跨域确定性服务质量保障。然而，此类基于时隙映射的跨域端到端方案涉及基于超周期（hyper-cycle）的跨网络域时间/频率同步，引入了大量的域间信令开销，系统复杂度较大。同时，目前也尚无成熟的方案实现跨域的时隙映射或时间/频率同步。其次，此类方案需要在时隙粒度上实现逐流逐跳的带宽资源分配。基于时隙的资源分配的本质是一个 NP 难的多维背包问题，且问题规模与超周期长度和链路数成指数关系[12]。在承载海量异构业务的端到端跨域场景中，不同的业务流会导致对应的超周期长度增加。此外，由于存在海量的骨干网链路，问题的解决空间也变得较大，使得问题的求解变得更加困难[4]。由于计算复杂度较高，一些基于列生成向量[13]、分支定界[12]、可满足性模型理论(Satisfiability Modulo Theories, SMT)[1]的资源分配算法无法直接应用。最后，时间敏感业务流可以划分为周期确定业务流、非周期确定业务流、和不确定业务流[1]。目前的跨域网络方案以超周期作为资源分配的窗口，仅能保证周期确定业务流的有界服务质量保证[4][10][11]。如何为多种类型的时间敏感业务，如非周期确定业务流、不确定业务流等，提供确定的服务质量保证仍需进一步研究[4]。

为了解决上述问题，提出一种跨域解耦的端到端确定性网络传输方案。该方案的特点是网络域之间不需要进行跨域时间同步或频率同步。首先，在各域内部采用相应的域内确定性网络协议，保证域内的传输时延有界。在用户侧使用循环队列转发机制（cyclic queuing forwarding，CQF）[13]，在网络侧使用可扩展确定性转发机制（scalable deterministic forwarding，SDF）[14]。在网络边缘设计逐流的离散整形功能，通过带宽保障的贪婪转发和逐流网络资源分配，压缩业务流在后续链路上的最大突发，保证传输质量的稳定和可靠。由于各域之间没有进行跨域时钟同步，无法采用基于时隙映射的方法构建基于时隙映射的端到端跨域通道。因此，采用了基于网络演算的跨域时延分析方法：通过递归方式获取业务流在沿途各个边缘节点上的到达曲线；使用网络演算的时分复用模型建立整形节点的服务曲线；通过最小加卷积运算获得整形节点的整形时延；最终建立业务流端到端跨域网络的传输。最后，提出一个基于拍卖模型的低复杂度在线业务流调度算法，在满足业务流端到端时延约束的情况下，最大化网络效率，提升网络吞吐量，减少带宽资源的租赁开销。由于不再以超周期作为资源分配窗口，所提端到端传输方案和调度机制适用于周期确定业务流、非周期确定业务流、和不确定业务流。

## 2 端到端跨域确定性网络设计

### 2.1 端到端跨域的确定性网络

构建如图 1 所示的端到端跨域确定性网络。图中包含 $N$ 个网络域，表示为集合 $\mathcal{N}$。从网络域的角度看，端到端跨域的确定性网络涉及 3 个部分，包括用户侧、网络侧、跨网络域[5]。

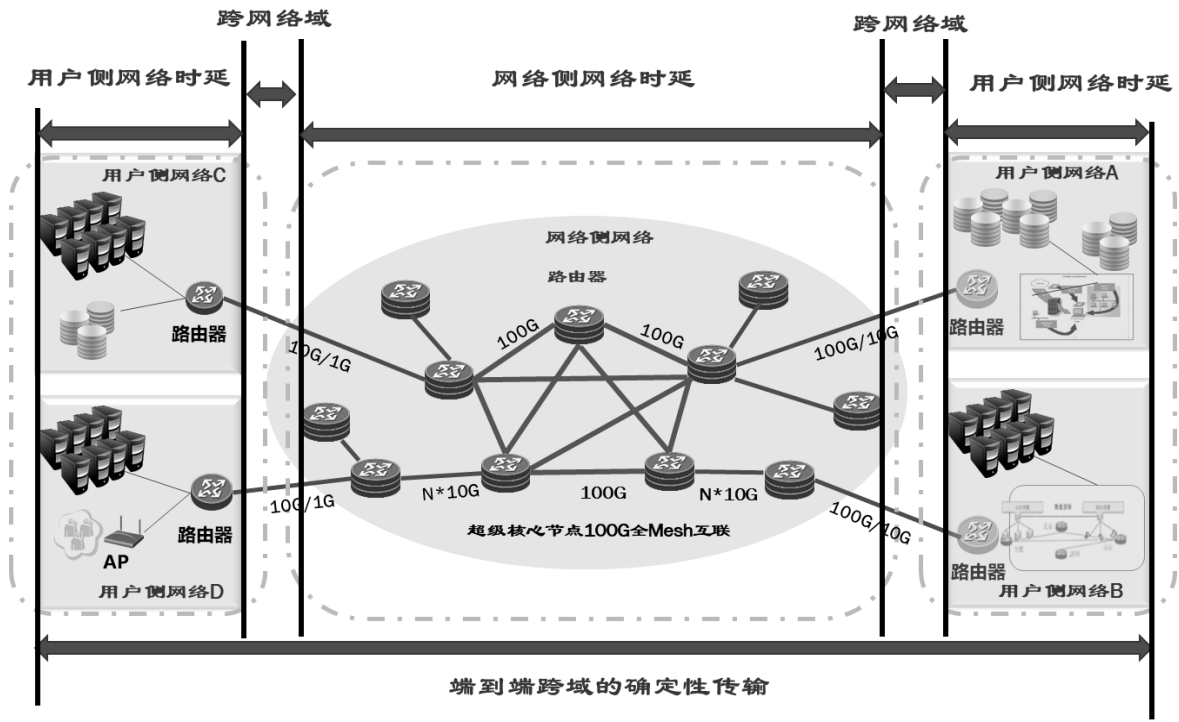

图 1 端到端跨域确定性网络

### 2.1.1 用户侧

用户侧主要指接入网,如企业网络、工厂园区网络等。源节点通过用户侧接入广域网络,与对端节点(目的节点)进行通信。以工业互联网为例,企业内网通常可以分为信息网络和生产网络。工业端设备,如机械臂等,作为源节点,通过生产网络和广域网,与云化的逻辑控制器(PLC,即为目的节点)进行远程通信。为实现高精度工业生产,源节点和目的节点之间需要建立具有时延上限的通信通道。控制/监控指令的传输不能丢包,传输时延/抖动不能超过给定门限。

通常,用户侧距离用户站点较近,其网络跳数一般小于 7 跳。以工业互联网为例,生产网络的覆盖范围通常为一个车间或一个园区。传统的生产网络采用的总线或工业以太网技术不能适应未来先进制造的要求。因此,可以采用二层确定性机制,如 TSN 网络技术,实现确定性接入。本文在用户侧统一采用 CQF 转发机制,实现确定性时延保障。

### 2.1.2 网络侧

网络侧主要指运营商网络,包括城域网、骨干网。网络侧是实现远距离确定性通信的关键。在工业互联网的应用场景中,需要实现 100 公里左右的远程控制,这种覆盖范围无法仅依靠用户侧网络实现。在消费类互联网场景中,如元宇宙,用户站点和服务节点数量巨大,仅依靠用户侧网络技术(如 TSN 技术)是不够的。为了在端到端场景中实现远距离确定性传输,网络侧网络通常采用 IP 协议,使用三层确定性协议 SDF 机制。

### 2.1.3 跨网络域

CQF 机制(用户侧)和 SDF 机制(网络侧)可以保障域内传输时延具有明确的上下界限。然而,要实现端到端的确定性传输,需要考虑如何抑制跨域过程中引入的不确定性。跨域情形包括用户侧网络跨越至运营商网络(网络侧网络)、跨越同一个运营商网络的多个网络、以及跨越不同运营商的网络。例如,用户可通过某一用户侧网络接入运营商网络,并试图访问其它运营商网络中的服务。为了模型的统一,将目的节点数据发送至用户侧网路也称为跨域(从目的节点到用户侧网络)。

实现端到端跨域确定性传输的关键之一在于如何在跨域网络节点上实施整形操作,消除后续传输中出现的微突发现象,降低排队时延的不确定性。为此,在各网络域入口处部署基于离散整形的跨域整形节点,并进行逐业务流的带宽预留,限制业务流的突发量。

### 2.2 网络与业务流模型

使用 $\mathcal{G} = \{\mathcal{V}, \mathcal{E}\} = \{(\mathcal{V}_n, \mathcal{E}_n)\}_{n \in \mathcal{N}} \bigcup \{e_{n,n'}^{cd}\}_{n,n' \in \mathcal{N}}$ 表示端到端跨域确定性网络。其中,$\mathcal{V}_n$ 为域 $n$ 中的节点(支持确定性功能的路由器/交换机),$\mathcal{E}_n$ 为域 $n$ 中的链路,$e_{n,n'}^{cd}$ 为连接域 $n$ 和域 $n'$ 的链路。域 $n$ 中链路带宽为 $\text{BW}_n$,跨域链路 $e_{n,n'}^{cd}$ 的带宽使用 $\text{BW}_{n,n'}$ 表示。用六元组表示业务流 $i \in \mathcal{I}$,记为 $\langle r_{i,0}, b_{i,0}, v_i^{\text{src}}, v_i^{\text{sink}}, \Gamma_i, v_i \rangle$。其中,$b_{i,0}$ 为业务流 $i \in \mathcal{I}$ 的最大突发,$r_{i,0}$ 为平均达到速率[15]。对于任意时

刻 $t$，业务流 $i \in \mathcal{I}$ 可以用经典的累积到达曲线进行刻画：

$$\alpha_{i,0}(t) = r_{i,0}t + b_{i,0}, \quad (1)$$

符号中的下标 $i$ 为业务流的索引，下标中的 0 表示业务流的初始参数；$v_i^{src}$ 和 $v_i^{sink}$ 分别为业务流 $i \in \mathcal{I}$ 的源节点和目的节点；$\Gamma_i$ 是业务流 $i \in \mathcal{I}$ 允许的最大端到端传输时延；$v_i$ 则给出了业务流的相对重要性，$v_i$ 越大说明业务流 $i \in \mathcal{I}$ 越重要。

使用 $\mathcal{P}_i$ 表示业务流 $i \in \mathcal{I}$ 可行的端到端路径集合，即任一集合元素 $p \in \mathcal{P}_i$ 是一条从节点 $v_i^{src}$ 到节点 $v_i^{sink}$ 的无环路径。令 $K_p$ 表示端到端路径 $p$ 涉及的网域个数，$n_{p,k}$ 表示路径 $p$ 沿路的第 $k$ 个网络域。根据网络域划分，端到端路径 $p$ 可被分解为多个分段，即 $p = \bigcup_{\forall k} p_k \cup_{\forall k} e_{n_{p,k-1}, n_{p,k}}^{cd}$。其中，$p_k$ 表示端到端路径 $p$ 在沿路第 $k$ 个域中的分段，$e_{n_{p,k-1},n_{p,k}}^{cd}$ 表示连接第 $k-1$ 个域和第 $k$ 个域的链路。根据定义，分段 $p_1$ 的起始位置为节点 $v_i^{src}$；分段 $p_{K_p}$ 的结束位置为节点 $v_i^{sink}$。网络模型与业务流模型的符号含义整理如表 1 所示。

## 2.3 基于贪婪转发的跨域离散整形

离散整形采用循环队列转发框架。在每个域 $n$ 的入口部署整形节点。为统一表述，所有业务流的源节点也视为网络域的边缘节点，在其出端口处部署对应的整形功能。整形节点将时间划分为长度为 $\tau_n$ 的时隙。若整形节点属于运营商网络，则与域内其它转发节点实现频率同步，$\tau_n$ 的取值与 SDF 的时隙长度相同。若整形节点属于用户侧网络，则与域内其它转发节点实现时钟同步，$\tau_n$ 的取值与 CQF 的时间片长度相同。注意到，整形节点不需要与相邻网域建立任何时间映射关系，各网络域的时钟是独立的。

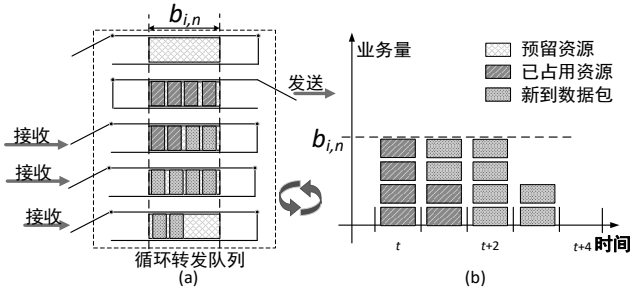

图 2 离散整形机制

如图 2(a)所示，在整形节点的每个出端口处，安装多个循环转发队列，不同转发队列对应各自的时隙。每个转发队列在对应时隙的起始时刻进行数据转发，直至对应时隙结束。假设整形节点的每个出口放置 $M$ 个循环转发队列，则时刻 $t$ 时，处于发送状态的队列索引为 $\mathrm{mod}(\lfloor t/\tau_n \rfloor, M)$。

业务流到达整形节点时，根据业务流在转发队列的资源占用情况，将业务数据转发到一个或多个队列中去。每个队列为不同的业务流预留容量 $b_{i,n}$（即整型参数，下标 $i$ 为业务流的索引）。在入队过程中，数据包将优先进入临近时隙的转发队列：假设当前时刻为 $t$，则依次检查索引为 $\mathrm{mod}(\lfloor t/\tau_n \rfloor, M)+1$ 至 $\mathrm{mod}(\lfloor t/\tau_n \rfloor, M)+M-1$ 的转发队列。若队列中业务流对应的预留容量 $b_{i,n}$ 尚未耗尽，则将数据包放入该队列；否则检查下一循环转发队列，直至数据包全部入队。

表 1 网络与业务流模型符号含义

| 符号 | 定义 |
| --- | --- |
| $\mathcal{V}$，$\mathcal{E}$，$\mathcal{V}_n$，$\mathcal{E}_n$，$e_{n,n'}^{cd}$ | 节点，边，域 $n$ 中的节点、链路连接域 $n$ 和域 $n'$ 的跨域链路 |
| $\mathrm{BW}_n$，$\mathrm{BW}_{n,n'}$ | 域 $n$ 中链路的带宽，连接域 $n$ 和域 $n'$ 跨域链路的带宽 |
| $r_{i,0}, b_{i,0}, v_i^{src}, v_i^{sink}, \Gamma_i, v_i$ | 业务流 $i \in \mathcal{I}$ 的平均速率，最大突发，源节点、目标节点、最大容忍时延、重要性因子 |
| $p$，$\mathcal{P}_i$，$K_p$，$n_{p,k}$，$p_k$，$e_{p,k,j}$ | 端到端路径，业务流 $i$ 的可行路径，路径 $p$ 经过的网络域个数，路径 $p$ 经过的第 $k$ 个网络域，路径 $p$ 在域 $n_{p,k}$ 中的分段，路径 $p$ 第 $k$ 个网络域的第 $j$ 跳链路 |
| $\tau_n$，$\tau_e$，$\tau_{p,k}^{trans}$，$\tau_{i,p,k}^{shaper}$，$\tau_i^{total}$ | 网络域 $n$ 中设备循环转发的周期长度，链路 $e$ 传播时延，网络域 $n_{p,k}$ 中使用路径 $p$ 的传输时延，业务流 $i$ 在网络域 $n_{p,k}$ 边缘的整形时延，业务流 $i$ 的端到端时延 |
| $p^{(i)}$，$\mathcal{B}_i$，$b_{i,n}$，$B_n^j$ | 业务流 $i$ 关联的传输路径，业务流 $i$ 关联的整形参数，业务流 $i$ 在网络域 $n$ 的整形参数，网络域 $n$ 可选的第 $j$ 个整形参数 |
| $M$，$m$，$N_{path}$，$N$ | 整形节点的循环队列数量，可选的整形参数个数，业务流最大的可行路径跳数，网络域的数量 |
| $\varepsilon_e$ | 链路 $e$ 的租赁开销 |
| $\mathbf{x} = \{x_i, \forall i \in \mathcal{I}\}$ | 所有业务流的准入控制变量 |
| $\mathbf{P} = \{p^{(i)}, \forall i \in \mathcal{I}\}$ | 所有业务流的路径选择 |
| $\mathcal{B}_i = \{b_{i,n_{p^{(i)},k}}, \forall k\}$ | 所有业务流的整形参数 |

图 2(b)展示了离散整形的一个例子，其中为业务流 $i$ 预留的容量 $b_{i,n} = 4k$。业务流 $i$ 在时隙 $t$ 内共到达 8 个数据包，每个数据包长度为 1k。此时，业务流 $i$ 在时隙 $t+1$ 对应队列中已有积压数据 2k。因此，新到达的 8k 数据将分别转发至时隙 $t+1$、$t+2$ 和 $t+3$ 对应的队列。每个转发队列中业务流 $i$ 的数据包数量不超过预留容量 $b_{i,n} = 4k$。

通过为每条时间敏感业务流分配不同的预留容量 $b_{i,n}$，可以有效限制业务流在本网络域的最大突

发，防止后续域内链路的拥塞。使用跨域离散整形的好处包括：

（1）**扩大了可调度业务流的种类**。采用经典的累积到达曲线描述业务流，使所提机制适用于周期确定业务流、非周期确定业务流和非确定业务流。传统的基于时隙跨域映射的方法（如文献[10]-[11]）仅能处理周期确定业务流的场景；

（2）**提高了超低延时场景的性能**。本文提出的跨域离散整数机制实质上采用了带宽约束的贪婪转发。在带宽资源充足的情况下，允许数据包直接进入下一个时隙对应的发送队列，缓存时间较短。传统的基于时隙跨域映射的方法需要根据时隙映射函数进行入队操作，需要考虑发送和接收时隙映射的最差情况，因此引入的跨域时间可能较长；

（3）**无需进行跨域时钟/频率同步**。在传输过程中，业务流的端到端时延可以分解为跨域时延（即整数时延加跨域链路传播时延）和域内传输时延。与基于跨域时隙映射的端到端确定性网络传输方案相比，主要差异在于跨域部分。离散整形节点采用带宽保障的贪婪转发方式，其跨域转发过程可以通过网络演算进行分析：预留带宽即为整形节点在每个时隙/时间片内提供的服务能力[16]。通过网络演算的相关算子，可以在不进行跨域时钟/频率同步的基础上，建立跨域时延上限。

### 2.4 网络域内传输时延计算方法

若业务流 $i \in \mathcal{I}$ 采用路径 $p$ 进行传输，则沿路径 $p$ 的第 $k$ 个网络域 $n_{p,k}$ 中的分段可表示为 $p_k = \{e_{p,k,1}, \cdots, e_{p,k,j}, \cdots, e_{p,k,|p_k|}\}$。其中，$e_{p,k,j}$ 为业务流 $i$ 在网络域 $n_{p,k}$ 中的第 $j$ 跳链路。网络域内部采用确定性网络传输机制，例如 CQF 和 SDF。

若网络域 $n_{p,k}$ 属于用户侧，则每跳传输的最大时延为 $\tau_{n_{p,k}}$。此时，域内传输时延的上界为：

$$\tau_{p,k}^{\text{trans}} = (|p_k|+1) \cdot \tau_{n_{p,k}}, \quad (2)$$

其中，$\tau_{n_{p,k}}$ 为网络域 $n_{p,k}$ 中 CQF 的时间片长度，$|p_k|$ 为路径 $p_k$ 的跳数[17]。

若网络域 $n_{p,k}$ 属于网络侧。此时，域内传输时延的上界可表示为：

$$\tau_{p,k}^{\text{trans}} = \sum_{j=1}^{|p_k|} \left( \tau_{e_{p,k,j}} + 2\tau_{n_{p,k}} \right) + \tau_{n_{p,k}}, \quad (3)$$

其中，$\tau_{n_{p,k}}$ 为 SDF 的时隙长度[18]。

### 2.5 域间跨域时延计算方法

本节采用网络演算计算跨域时延上界。根据网络演算方法，跨域时延可以根据整形节点的到达曲线和服务曲线获得。

整形节点为业务流 $i \in \mathcal{I}$ 提供的服务曲线可以根据 TDM 的经典服务模型获得。若业务流 $i \in \mathcal{I}$ 采用路径 $p$ 进行传输，则沿路径第 $k$ 个网络域为网络域 $n_{p,k}$。网络域 $n_{p,k}$ 边缘整形节点为业务流 $i$ 预留的容量为 $b_{i,n_{p,k}}$，则整形节点对业务流 $i \in \mathcal{I}$ 的服务曲线可以表示为[16]：

$$\beta_{i,p,k}(t) = \left\lceil \frac{b_{i,n_{p,k}}}{\tau_{n_{p,k}}} \right\rceil \cdot \max\left\{0, t - 2\tau_{n_{p,k}}\right\}, \quad (4)$$

其中，$\tau_{n_{p,k}}$ 为域 $n_{p,k}$ 中节点循环转发的时隙长度。若网络域 $n_{p,k}$ 为用户侧，则 $\tau_{n_{p,k}}$ 为网络域 $n_{p,k}$ 中 CQF 的时间片长度；否则，$\tau_{n_{p,k}}$ 为网络域 $n_{p,k}$ 中 SDF 的时隙长度。假设业务流 $i$ 在网络域 $n_{p,k}$ 边缘的到达曲线为 $\alpha_{i,p,k}(t)$，则业务流 $i$ 的整形时延的上界可表示为

$$\tau_{i,p,k}^{\text{shaper}} = \sup\left\{ \alpha_{i,p,k}(t) - \alpha_{i,p,k}(t) \otimes \beta_{i,p,k}(t) \right\}, \quad (5)$$

其中，$\beta_{i,p,k}(t)$ 为公式(2)中的服务曲线，$\otimes$ 为最小加卷积算符[15]。

## 3 时延有界的端到端跨域调度模型

### 3.1 决策变量

主要考虑下列决策变量：

**准入控制变量**：$\mathbf{x} = \{x_i, \forall i \in \mathcal{I}\}$，其中 $x_i = 1$ 代表允许业务 $i \in \mathcal{I}$ 接入传输；否则，$x_i = 0$。网络需要根据现网内的资源剩余情况和业务时延约束要求，判断是否允许业务流进行接入。若允许业务流接入（即 $x_i = 1$），则需为业务流分配传输资源，包括给定传输路径和整形参数。

**传输路径**：使用 $p^i = \{p_k^i\}_{\forall k} \cup \{e_{n_{p^i,k-1}, n_{p^i,k}}^{\text{cd}}\}_{\forall k} \in \mathcal{P}_i$ 表示业务 $i \in \mathcal{I}$ 关联的端到端路径，其中 $p_k^i$ 表示业务流 $i \in \mathcal{I}$ 在第 $k$ 个网络域中的分段。若 $x_i = 0$，则 $p^i = \varnothing$；否则，$p^i \neq \varnothing$。令 $\mathbf{P} = \{p^i\}_{\forall i \in \mathcal{I}}$ 表示所有业务流的路径选择。

**整形参数**：$\mathcal{B}_i = \{b_{i,n_{p^i,k}}, \forall k\}$，其中 $b_{i,n_{p^i,k}}$ 为业务流 $i \in \mathcal{I}$ 在第 $k$ 个网络域 $n_{p^{(i)},k}$ 边缘整形节点的预留容量。令 $\mathbf{B} = \{\mathcal{B}_i\}_{\forall i \in \mathcal{I}}$ 表示所有业务流整形参数的选择。

## 3.2 约束条件
### 3.2.1 时延约束

对于已准入业务流 $i \in \mathcal{I}$，假设给定其传输路径和整形参数，包括 $p$ 和 $\mathcal{B}_i$。根据不同的网络域，业务流 $i \in \mathcal{I}$ 端到端的时延上界可分解为：

$$\tau_i^{\text{total}} = \sum_{k=1}^{K_p}\left(\tau_{i,p,k}^{\text{shaper}} + \tau_{p,k}^{\text{trans}}\right) + \sum_{k=2}^{K_p} \tau_{e_{n_{p,k-1},n_{p,k}}^{cd}}, \quad (6)$$

其中，$\tau_{e_{n_{p,k-1},n_{p,k}}^{cd}}$ 为业务流从第 $k-1$ 个网域 $n_{p,k-1}$ 流入到第 $k$ 个网域 $n_{p,k}$ 的链路传播时延，$\tau_{i,p,k}^{\text{shaper}}$ 为第 $k$ 个网域的整形时延，$\tau_{p,k}^{\text{trans}}$ 为第 $k$ 个网域中的传输时延。

端到端时延公式(6)中，延时项 $\tau_{e_{n_{p,k-1},n_{p,k}}^{cd}}$ 和 $\tau_{p,k}^{\text{trans}}$ 的上界可以通过路径 $p$ 直接获取：域间链路时延 $\tau_{e_{n_{p,k-1},n_{p,k}}^{cd}}$ 通过测量链路 $e_{n_{p,k-1},n_{p,k}}^{cd}$ 的传播时延得到；域内传输时延 $\tau_{p,k}^{\text{trans}}$ 的上界取决于路径 $p_k$ 和域内时间片（或时隙）长度 $\tau_{n_{p,k}}$，具体见公式(4)和公式(5)。整形时延 $\tau_{i,p,k}^{\text{shaper}}$ 的上界取决于业务流 $i \in \mathcal{I}$ 在网域 $n_{p,k}$ 的到达曲线和对应的整形参数 $b_{i,n_{p,k}}$。下面，具体阐述如何计算业务流 $i \in \mathcal{I}$ 在网域 $n_{p,k}$ 边缘的整形时延。

业务流 $i \in \mathcal{I}$ 达到路径上第 $k$ 个网络域 $n_{p,k}$ 边缘的整形节点。当 $k=1$ 时，到达曲线表示为[16]：

$$\alpha_{i,p,1}(t) = r_{i,0} \cdot t + b_{i,0}, \quad (7)$$

当 $k>1$ 时，到达曲线可表示为[16]：

$$\alpha_{i,p,k}(t) = \left(b_{i,n_{p,k-1}}\left\lceil\frac{\tau_{n_{p,k-1}}}{\tau_{n_{p,k}}}\right\rceil\left\lceil\frac{t+\tau_{n_{p,k}}}{\tau_{n_{p,k}}}\right\rceil\right) \otimes \left(\text{BW}_{n_{p,k-1},n_{p,k}} \cdot t\right), (8)$$

其中，$b_{i,n_{p,k-1}}\lceil \tau_{n_{p,k-1}}/\tau_{n_{p,k}} \rceil$ 为每个时隙中业务流 $i$ 到达的比特数目。可以看出，在第 $k$ 个网络域边缘（$k>1$ 时），业务流 $i \in \mathcal{I}$ 的到达曲线取决于上一个网络域整形参数 $b_{i,n_{p,k-1}}$ 与链路 $e_{n_{p,k-1},n_{p,k}}^{cd}$ 的带宽。

综上所述，第 $k$ 个网络域中，业务流 $i$ 的到达曲线可以表示为：

$$\alpha_{i,p,k}(t) = \begin{cases} r_{i,0} \cdot t + b_{i,0} & k=1 \\ \left(b_{i,n_{p,k-1}}\left\lceil\frac{\tau_{n_{p,k-1}}}{\tau_{n_{p,k}}}\right\rceil\left\lceil\frac{t+\tau_{n_{p,k}}}{\tau_{n_{p,k}}}\right\rceil\right) \otimes \left(\text{BW}_{n_{p,k-1},n_{p,k}} \cdot t\right) & k \geq 2 \end{cases}.$$

由于服务曲线已知，通过到达曲线可获得第 $k$ 个网络域的整形时延上界：

$$\tau_{i,p,k}^{\text{shaper}} = \sup\left|\alpha_{i,p,k}(t) - \alpha_{i,p,k}(t) \otimes \beta_{i,p,k}(t)\right|, \quad (9)$$

其中，$\beta_{i,p,k}(t)$ 为第 $k$ 个网络域整形节点的服务曲线。

根据式(6)，业务流端到端时延 $\Gamma_i$ 的约束可以表示为：

$$\sum_k \left(\tau_{e_{n_{p,k-1},n_{p,k}}^{cd}} + \tau_{i,p,k}^{\text{shaper}} + \tau_{p,k}^{\text{trans}}\right) \leq \Gamma_i. \quad (10)$$

约束（10）对所有业务流 $i \in \{i \mid x_i=1, \forall i \in \mathcal{I}\}$ 成立。

### 3.2.2 资源约束

已准入的业务流 $i \in \{x_i=1 \mid i \in \mathcal{I}\}$ 会消耗其传输路径上的带宽资源。假设对所有准入业务流 $i \in \{x_i=1 \mid i \in \mathcal{I}\}$，给定其传输路径和整形参数，包括 $p^i$ 和 $\mathcal{B}_i$。

对于任意链路 $e \in \mathcal{E}$，令 $\chi_{e,n}(p)$ 表示路径 $p$ 是否包含链路 $e$。即若 $e \in p$ 且链路 $e$ 属于网络域 $n$，则 $\chi_{e,n}(p)=1$，否则 $\chi_{e,n}(p)=0$。链路 $e \in \mathcal{E}$ 的带宽资源约束可表示为：

$$w_e = \sum_{i \in \mathcal{I}}\sum_k \left[b_{i,p^{(i)},k} \cdot \chi_{e,n_{p^{(i)},k}}\left(p^{(i)}\right)\right] \leq \text{BW}_{n(e)} \cdot \tau_{n(e)}. (11)$$

其中，$n(e)$ 表示链路 $e$ 所在的网络域。

## 3.3 目标函数

考虑在端到端确定性网络中，最大化网络效用，即在满足时延约束(10)和资源约束(11)情况下，最大化确定性业务流接入的加权和减去确定性业务流预留带宽的租赁开销：

$$F(\mathbf{x},\mathbf{P},\mathbf{B}) = \sum_{i \in \mathcal{I}} v_i x_i - \sum_{e \in \mathcal{E}} \varepsilon_e w_e, \quad (12)$$

其中，$v_i$ 表示业务流 $i \in \mathcal{I}$ 的相对重要程度，$v_i$ 越大，业务流 $i \in \mathcal{I}$ 越重要。$w_e$ 是链路 $e$ 预留给所有确定性业务流的资源。$\varepsilon_e$ 是链路 $e$ 上单位带宽资源的租赁价格。

目标函数 $F(\mathbf{x},\mathbf{P},\mathbf{B})$ 由两部分组成。第一部分 $\sum_{i \in \mathcal{I}} v_i x_i$ 是所有业务流准入变量 $x_i$ 的加权和，权重为 $v_i$，代表不同业务流的重要性。$\sum_{i \in \mathcal{I}} v_i x_i$ 越大，说明网络接入业务流的质量越高，数目越大。第二部分 $\sum_{e \in \mathcal{E}} \varepsilon_e w_e$ 是网络带宽的租赁开销。最大化目标函数 $F(\mathbf{x},\mathbf{P},\mathbf{B})$ 的目的是在准入尽可能多、尽可能重要的业务流的同时，降低网络带宽的租赁开销。综上，考虑的优化问题可以总结为：

$$\max_{\mathbf{x}, \mathbf{P}=(p^{(i)}, \forall i \in \mathcal{I}), \mathbf{B}=(\mathcal{B}_i, \forall i \in \mathcal{I})} \sum_{i \in \mathcal{I}} v_i x_i - \sum_{e \in \mathcal{E}} \varepsilon_e w_e$$

$$s.t. \quad C1: x_i = \{0,1\}, \forall i \in \mathcal{I}$$
$$C2: p^{(i)} \in \mathcal{P}_i, \forall i \in \mathcal{I} \quad , \quad (13)$$
$$C3: b_{i,n} \in \{B_n^1, \cdots, B_n^m\}, \forall i \in \mathcal{I}, \forall n$$
$$\text{constraint equations (10),(11).}$$

其中，C1，C2 和 C3 给出了 $\mathbf{x}$，$p^i$ 和 $\mathcal{B}_i$ 对应物理含义的约束。对于 $b_{i,n}$ 的取值，本文假设其仅可以取 $m$ 个离散的值，包括 $B_n^1, \cdots, B_n^m$。

## 4 基于组合拍卖的在线调度算法

典型的跨域场景包括远程工业控制、高清视频传输、远程 AR/VR 交互、远程手术等。从网络的角度来看，这些业务流是序贯到达的。因此，问题(13)需要进行在线求解[1]。当业务流 $i$ 接入网络之前，调度算法需要为其分配网络资源，包括确定传输路径 $p^i$ 和整形参数 $\mathcal{B}_i$。若网络资源足够，则调度算法下发对应 $p^i$ 和 $\mathcal{B}_i$，并告知业务流 $i$ 可以进行业务传输；否则，拒绝业务流 $i$ 的接入请求。

---

**算法 1** 在线业务流调度算法

1 For 业务流 $i \in \mathcal{I}$ 到达时 do // 迭代
2   业务流 $i \in \mathcal{I}$ 向网络提交相关参数，包括 $r_{i,0}$, $b_{i,0}$, $v_i^{\text{src}}$, $v_i^{\text{sink}}$, $\Gamma_i$ 和 $v_i$。
3   网络根据 $v_i^{\text{src}}$ 和 $v_i^{\text{sink}}$，使用**算法 2** 枚举所有可能的资源组合 $\bar{\pi}_i$。
4   网络筛选出满足约束条件(10)的可行资源组合 $\pi_i$。
5   网络根据资源定价函数(16)，计算所有可行资源组合 $(p, \mathcal{B}_i(p)) \in \pi_i$ 的价格。
6   令 $(p^*, \mathcal{B}_i^*(p^*)) = \arg\min_{(p, \mathcal{B}_i(p)) \in \pi_i} \{c(p, \mathcal{B}_i(p))\}$ 和 $c_{\min} = \min_{(p, \mathcal{B}_i(p)) \in \pi_i} \{c(p, \mathcal{B}_i(p))\}$
7   If $c_{\min} \leq v_i$ 且 $(p^*, \mathcal{B}_i^*(p^*))$ 满足约束(11) do
8     接入业务流 $i \in \mathcal{I}$
9     业务流 $i \in \mathcal{I}$ 传输路径为 $p^{(i)} = p^*$，整型参数为 $\mathcal{B}_i = \mathcal{B}_i^*(p^*)$
10   else
11     拒绝业务流 $i \in \mathcal{I}$
12   End
13 End

---

提出了基于组合拍卖的在线调度算法，实现问题(13)的在线求解。网络将链路带宽资源视为拍卖商品（即 items）。业务流 $i$ 作为竞拍者向网络购买链路带宽资源的组合，实现端到端跨域的确定性传输。为实现端到端的确定性传输，每个资源组合可以使用二元组 $(p, \mathcal{B}_i(p))$ 进行唯一标识，其中 $p \in \mathcal{P}_i$ 表示端到端传输路径；$\mathcal{B}_i(p)$ 为路径沿路的整形参数。

在线分配算法如算法 1 所示。当业务流 $i$ 到达时，执行 Line 2-12。首先，业务流将向网络发起资源购买请求（Line 2），向网络提交业务流的相关信息，包括（1）达到信息平均速率和突发；（2）业务的源节点和目标节点；（3）业务流的端到端时延要求 $\Gamma_i$。同时，向网络提交业务流的竞拍价格（即 bids，其值等于业务流参数 $v_i$）。

网络收到请求后，使用算法 2 枚举所有可能的资源分配组合 $\bar{\pi}_i$（见算法 1 的 Line 3）。在算法 2 中，网络使用遍历算法获取所有从节点 $v_i^{\text{src}}$ 到节点 $v_i^{\text{sink}}$ 的可行路径集合 $\mathcal{P}_i$。每条端到端可行路径 $p \in \mathcal{P}_i$ 对应 $K_p$ 个整形参数 $\mathcal{B}_i(p) = \{b_{i, n_{p,k}}\}_{k=1}^{K_p}$。其中，$b_{i, n_{p,k}}$ 可以取 $m$ 个离散值，包括 $\{B_{n_{p,k}}^1, \cdots, B_{n_{p,k}}^m\}$。当给定一条端到端可行路径 $p \in \mathcal{P}_i$ 时，可行整形参数组合 $\mathcal{B}_i(p)$ 有 $m^{K_p}$ 个。因此，可行资源组合的集合 $\bar{\pi}_i$ 中约束的个数最多为 $|\mathcal{P}_i| m^N$。

---

**算法 2** 业务流错误!不能通过编辑域代码创建对象。的资源错误!不能通过编辑域代码创建对象。枚举方法

1 网络根据业务流 $i \in \mathcal{I}$ 的 $v_i^{\text{src}}$ 和 $v_i^{\text{sink}}$，获取路径集合 $\mathcal{P}_i$
2 令 $\bar{\pi}_i = \varnothing$
3 For any 路径 $p \in \mathcal{P}_i$ do
4   罗列路径 $p$ 上 $K_p$ 个域，每个域有 $m$ 个可用的整形参数。
5   对 $K_p$ 个整形参数的可行值进行排列组合，形成集合 $\Sigma_p$（$|\Sigma_p| = m^{K_p}$）；
6   For any 集合 $\Sigma_p$ 中的元素 $\mathcal{B}_i(p)$ do
7     令 $\bar{\pi}_i = \bar{\pi}_i \cup (p, \mathcal{B}_i(p))$
8   End
9 End

---

接着，网络对 $\bar{\pi}_i$ 中所有资源分配组合进行筛选，得到满足约束条件(10)的可行资源组合 $\pi_i$（见算法 1 的 Line 4）。

Line 5 中对资源的合理定价是在线调度算法性能优劣的关键。若资源分配过于"激进"，即对资源的定价过低，则网络资源将优先满足先到达的业务流。这可能导致后到的、重要性较大的业务流无

法获取资源，进而使整体调度性能下降。若资源分配过于"保守"，即对资源的定价过高，则部分重要性较低的业务流无法接入网络。这可能导致资源浪费，同样使整体调度性能下降[19]。为解决上述问题，定义链路资源定价函数：

$$\phi_e(\tilde{w}_e) = (v_{\min} - \varepsilon_e)\left(\frac{v_{\max} - \varepsilon_e}{v_{\min} - \varepsilon_e}\right)^{\frac{\tilde{w}_e}{BW_e}}, \quad (14)$$

其中，$\tilde{w}_e$ 是链路 $e$ 当前的已被分配出去的资源；$v_{\min}$ 和 $v_{\max}$ 分别定义为：

$$\begin{cases} v_{\min} = \min\left\{\dfrac{v_i}{r_{i,0}}, \forall i \in \mathcal{I}\right\} \\ v_{\max} = \max\left\{\dfrac{v_i}{r_{i,0}}, \forall i \in \mathcal{I}\right\} \end{cases}. \quad (15)$$

最终，资源分配组合 $(p, \mathcal{B}_i(p))$ 的价格可表示为：

$$c(p, \mathcal{B}_i(p)) = \sum_k \sum_e b_{i,n_{p,k}} \phi_e(w_e) \chi_{e,n_{p,k}}(p), \quad (16)$$

即资源组合 $(p, \mathcal{B}_i(p)) \in \pi_i$ 的价格是其所涉及链路资源的价格之和。计算集合 $\pi_i$ 中所有可行资源组合的价格 $\{c(p, \mathcal{B}_i(p))\}_{\forall (p, \mathcal{B}_i(p)) \in \pi_i}$，并获取价格最小的资源组合 $(p^*, \mathcal{B}_i^*(p^*))$ 作为候选调度策略，并记下对应的资源组合价格 $c_{\min}$。

算法 1 通过 Line 7 的 IF 条件判断是否准入业务流 $i$：若 $c_{\min} \leq v_i$ 且 $(p^*, \mathcal{B}_i^*(p^*))$ 满足约束(11)，则准入业务流 $i$；否则，拒绝业务流 $i$ 接入。若准入业务流 $i$，则将对应传输路径 $p^{(i)} = p^*$ 和整形参数 $\mathcal{B}_i = \mathcal{B}_i^*(p^*)$ 下发至各网络域转发设备。业务流 $i$ 根据分配的端到端路径及整形参数进行数据转发。

**复杂度分析**：每个业务流到达时，都会执行算法 1 的步骤 Line 2-12。其中，Line 3 需要枚举所有满足约束条件(10)的资源组合 $(p, \mathcal{B}_i(p)) \in \bar{\pi}_i$。在该步骤中，最多共有 $|\mathcal{P}_i|m^N$ 种不同的资源组合，穷举的复杂度为 $\mathcal{O}(|\mathcal{P}_i|m^N)$。算法 1 的步骤 Line 4 对资源组合进行最大时延约束的筛选，得到集合 $\pi_i$。根据 3.2.1 章节的计算方法，此步骤复杂度的上界为 $\mathcal{O}(N|\mathcal{P}_i|m^N)$。算法 1 的步骤 Line 5 和 Line 6 计算可行组合 $(p, \mathcal{B}_i(p)) \in \pi_i$ 的价格，每条路径上价格的计算至多涉及 $|\mathcal{E}|$ 条边。因此，此步骤复杂度的上界可表示为 $\mathcal{O}(|\mathcal{P}_i||\mathcal{E}|m^N)$。Line 7 中判断约束条件(11)是否成立的复杂度为 $\mathcal{O}(|\mathcal{E}|)$ 综上，完成所有业务流调度的时间复杂度为：

$$\mathcal{O}\left[N_{\text{path}}m^N(N+|\mathcal{E}|)+|\mathcal{E}|\right]|\mathcal{I}|, \quad (17)$$

其中，$N_{\text{path}} = \max_i |\mathcal{P}_i|$。

# 5 仿真实验

## 5.1 仿真环境

仿真环境基于 OMNeT++ 与 Matlab 构建，其中 OMNeT++ 负责数据面仿真，Matlab 负责生成在线调度方案。

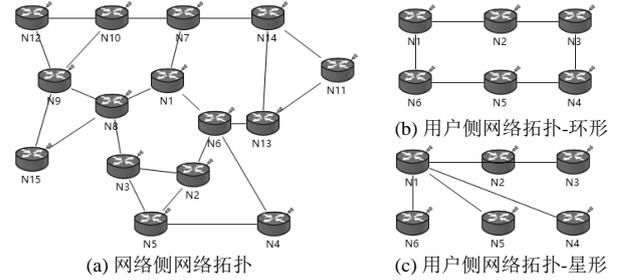

(a) 网络侧网络拓扑　　(b) 用户侧网络拓扑-环形

(c) 用户侧网络拓扑-星形

图 3 网络侧、用户侧仿真拓扑

表 2 仿真参数设置

| 参数 | 取值 |
| --- | --- |
| $\tau_n$ | 网络侧：$10\mu s$；用户侧：$2\mu s$ 或 $8\mu s$ |
| $\tau_e$ | 网络侧：$150\mu s$；用户侧：$0\mu s$ |
| $\varepsilon_e$ | 7.5 |
| $BW_n$ | 网络侧带宽：10Gbps；用户侧带宽：1Gbps； |
| $BW_{n,n'}$ | 1Gbps |
| $N$ | 6 |
| $|\mathcal{I}|$ | 500 |
| $\Gamma_i$ | 1ms |
| $v_i$ | 随机分布在 1-100 之间； |
| $b_{i,0}$ | 1500Byte |
| $r_{i,0}$ | 7.5Mbps（发送周期 $200\mu s$）；5Mbps（发送周期 $300\mu s$）；3.75Mbps（发送周期 $400\mu s$）；3Mbps（发送周期 $500\mu s$）； |
| $N_{\text{path}}$ | 10 |
| $M$ | 10 |
| $m$ | 3 （对应 $B_n^1 = 5\text{Mbps}$，$B_n^2 = 10\text{Mbps}$，$B_n^3 = 100\text{Mbps}$） |

如图 3(a)所示，基于亚特兰大城域网的真实拓扑构建仿真场景。仿真网络包含 1 个网络侧网络和

5 个用户侧网络（即 N=6）。网络侧网络包含 15 个节点，链路传播时延约为 $150 \mu s$（对应链路长度 30km 左右），链路带宽为 10Gbps。按照所提方案，每个节点运行 SDF 机制，时隙长度为 $10 \mu s$。5 个用户侧网络通过网络侧网络相互联通。如图 3 (b)所示，用户侧网络拓扑选用 TSN 网络常用的环形或星形拓扑，带宽设定为 1Gbps。用户侧网络设备运行 CQF 机制，时间片长度为 $2 \mu s$ 或 $8 \mu s$。用户侧网络通过各自的 N1 节点与网络侧网络联通。三个环形拓扑的用户侧网络分别通过 N1、N13、N3、N9 节点与网络侧网络连接，连接带宽为 10Gbps；两个线性拓扑的用户侧网络分别通过 N1、N14 节点与网络侧网络连接，连接带宽也为 10Gbps。网络中，所有链路带宽租赁费用 $\varepsilon_e$ 设为 7.5。

仿真生成 1500 个业务流，其中 500 个为时间敏感业务流。每个时间敏感业务流数据包的大小为 1500Bytes，最大传输时延为 1ms，目的节点为网络侧网络的 N1 节点，权重 $v_i$ 随机分布在 1-100 之间。每个用户侧网络负责接入 100 个时间敏感业务流。在传输过程中，业务流必定会通过至少 1 个用户侧网络和 1 个网络侧网络。

为了证明有效性，在仿真中将所提出的机制与传统"尽力而为"的转发网络进行比较。随后，选取具有代表性的、基于超周期的跨域机制[10]作为对比机制，突出所提传输机制在端到端时延保障方面与调度复杂度方面的优越性。最后，以在线贪心算法作为对比算法，检验所提调度算法的算法效率，仿真参数设置详见表 2。

## 5.2 仿真结果

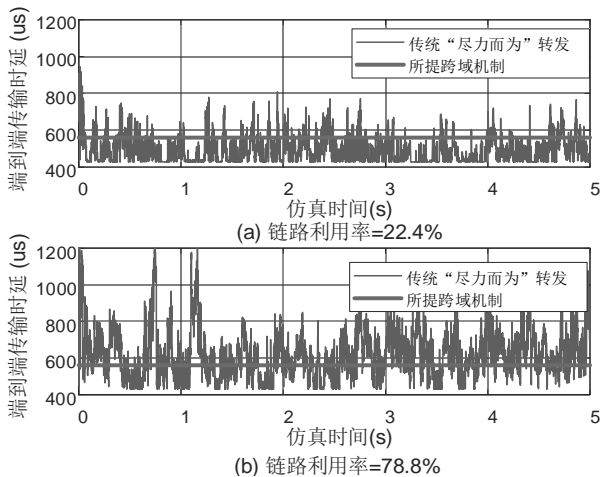

图 4 端到端确定性时延保障

首先，与传统"尽力而为"转发机制的端到端时延变化进行对比，验证所提机制在不同背景流负载下的端到端 QoS 保障能力。此处的背景流指的是权重为 0 的时间不敏感业务流，如网页浏览数据等。从图 4 中可以看出，所提机制可以有效实现时延上有界的端到端跨域确定性传输。由于微突发的原因，即使网络处于轻载情况（链路平均利用率为 22.4%），传统"尽力而为"转发方式在 5s 内的最大时延为 $985 \mu s$，最大抖动超过 $550 \mu s$。当链路负载增加至 78.4%时，传统"尽力而为"转发方式的时延抖动增大至 $700 \mu s$ 以上。另一方面，所提跨域机制可以有效的将端到端时延保证在 $600 \mu s$ 以下，且不受背景流负载的影响。由于抖动较小，所提机制时延曲线在图 4 中近似一条直线。同时值得注意，由于采用基于时分的循环队列转发，在传输过程中引入额外时延，导致所提机制的端到端时延大于尽力而为转发的最小值。

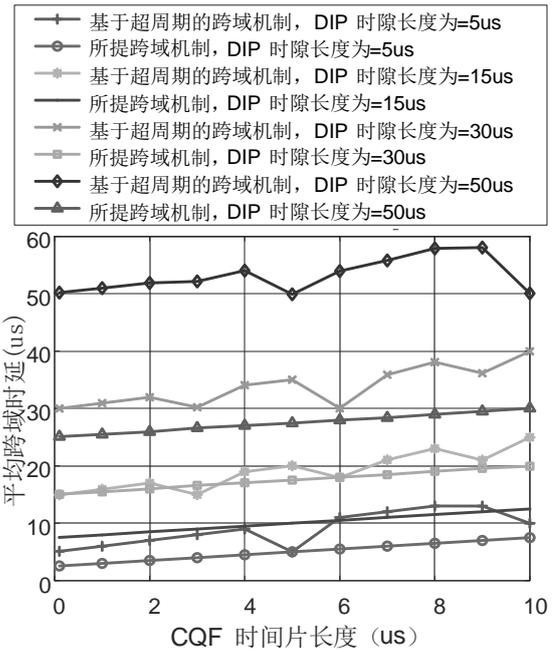

图 5 平均跨域传输时延

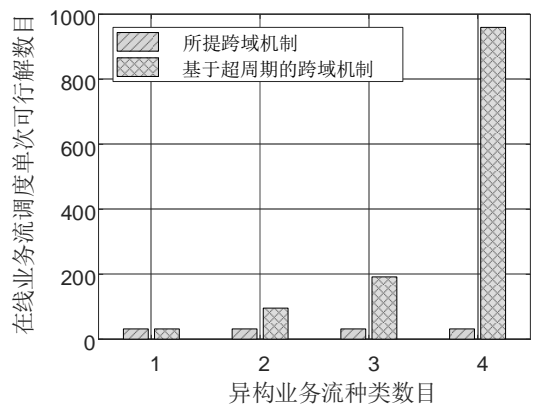

图 6 不同调度机制的在线调度复杂度

图 5 表明，与其它跨域确定性技术相比，所提机制可以实现更低的跨域传输时延（即整形时延+跨域链路时延）。由于所提机制与传统跨域机制在网络域内采用类似的确定性传输机制，端到端时延差异部分主要来源于跨域传输部分。传统的跨域机

制需要在网络边缘建立时隙映射。由于相邻域时间片/时隙长度有差异，时隙映射会将数据包映射到更靠后的时隙，以消弭因跨域导致的时隙偏差。当相邻两个域之间的时隙差异较大时，此阶段引入的时延较大。所提机制在边缘采用基于离散整形的先到先走策略，避开了因时隙对齐带来的额外时延。因此，与现有的确定性跨域机制相比，跨域时延被压缩了。

图 6 给出了在线调度单个业务流时，不同跨域机制可行解的数量。考虑异构业务流的发送周期为 $200\,\mu s$，$300\,\mu s$，$400\,\mu s$ 和 $500\,\mu s$。图中给出了发送周期为 $200\,\mu s$ 业务流的单次调度可行解数量。可以看出，所提机制有较好的可扩展性，即单次调度可行解的数量与业务流种类无关。而基于超周期的调度机制，可行解会随着业务流种类的变多，急剧变大。例如，传输周期为 $200\,\mu s$，$300\,\mu s$，$400\,\mu s$ 的 3 种时间敏感异构业务流共网传输，其超周期为 1.20ms。网络需要确定每个 1.20ms 时间段内，业务流所占用时隙的组合。通常来说，当网络中含有 10 种以上的异构时间敏感业务流，超周期的长度就非常长，业务流所占时隙组合数量非常大。另一方面，如第四章中的分析，所提调度机制的复杂度仅与网络规模、业务流数目、整形参数配置有关，与业务流的特性无关。因此，随着异构业务的加入，在线调度可行解数目不变，为常数。

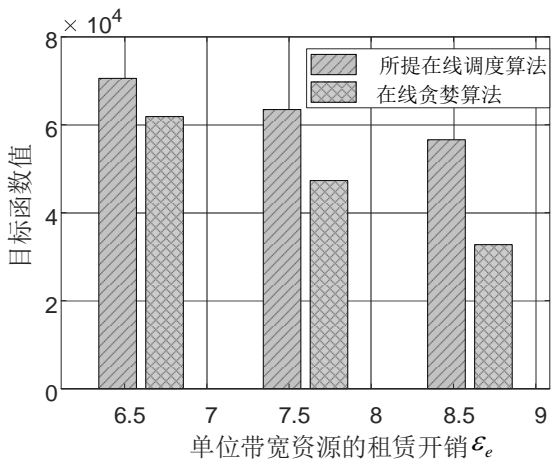

图 7 不同租赁开销下的算法效率

图 7 给出了不同租赁开销 $\varepsilon_e$ 下的算法效率。图中纵坐标代表目标函数 $F(\mathbf{x},\mathbf{P},\mathbf{B})$ 的值，数值约大说明算法优化性能越好。可以看出，与在线贪婪算法相比，所提机制可以实现较高的目标函数值。同时，随着租赁开销的进一步增大，所提机制所占优势逐渐放大。由于业务流是续贯到达的，网络无法判断未来是否有更加"重要"的业务流（权重更大）。若资源分配较为激进，将导致未来权重更大的业务流无法获取传输资源，降低目标函数值；反之亦然。所提机制通过定义一个单调递增的资源定价函数解决该问题，即资源的价格随着资源利用率的增加而单调递增。后到的业务若权重较大，则仍可购买相关资源，接入网络；若权重相对较小，网络将拒绝本次业务，为后续业务预留资源。

# 6 结束语

现有确定性网络技术基本解决了网络域内服务质量无法保障的问题。因此，为实现端到端的、跨多个网络域的确定性服务质量保障，需要着重解决域间传输引入的传输质量不确定问题。本文提出一个域间松耦合的端到端确定性传输与调度机制。该机制实现了相邻网络时间域的解耦，解决了现有方案部署及调度复杂程度高的问题。具体来说，在域入口处部署了逐流的跨域整形功能，通过带宽保障的逐流网络资源分配，降低业务流在后续链路上的最大突发带宽。同时，通过网络演算构建业务流跨域传输的时延上界。通过拍卖模型联合考虑每条时间敏感业务流的带宽分配和传输路径，实现了时间敏感业务流的低复杂的在线调度。最后，通过仿真验证了所提方法和算法的有效性。